\documentclass{article}
\usepackage{spconf,amsmath,graphicx,amsfonts}

\title{Cracking the cocktail party problem by multi-beam deep attractor network}
\name{Zhuo Chen, Jinyu Li, Xiong Xiao, Takuya Yoshioka, Huaming Wang, Zhenghao Wang, Yifan Gong\vspace{-.5em}}
\address{Microsoft AI and Research\vspace{-1em}}
\begin{document}
\ninept
\maketitle
\begin{abstract}

While recent progresses in neural network approaches to single-channel speech separation, or more generally  the cocktail party problem, achieved significant improvement, their performance for complex mixtures is still not satisfactory. In this work, we propose a novel multi-channel framework for multi-talker separation. In the proposed model, an input multi-channel mixture signal is firstly converted to a set of beamformed signals using fixed beam patterns. For this beamforming, we propose to use differential beamformers as they are more suitable for speech separation. Then each beamformed signal is fed into a single-channel anchored deep attractor network to generate separated signals. And the final separation is acquired by post selecting the separating output for each beams.  To evaluate the proposed system, we create a challenging dataset comprising mixtures of 2, 3 or 4 speakers. Our results show that the proposed system largely improves the state of the art in speech separation, achieving 11.5 dB, 11.76 dB and 11.02 dB average signal-to-distortion ratio improvement for  4, 3 and 2 overlapped speaker mixtures, which is comparable to the performance of a minimum variance distortionless response beamformer that uses oracle location, source, and noise information. We also run speech recognition with a clean trained acoustic model on the separated speech, achieving relative word error rate (WER) reduction of 45.76\%, 59.40\% and 62.80\% on fully overlapped speech of 4, 3 and 2 speakers, respectively. With a far talk acoustic model, the WER is further reduced.
\end{abstract}
\section{Introduction}
\label{sec:intro}








The cocktail party problem has been one of the most difficult challenges in audio signal processing for more than 60 years \cite{cherry1953some,ASB90,brown1994computational}. In this problem, the task is to separate and recognize each speaker in highly overlapped speech recordings, as frequently happens in a cocktail party. Although humans can solve this problem naturally without much effort, it is extremely difficult to build an effective system to model this process. Because of the large variation in mixing sources, the cocktail party problem remains unsolved.


Prior to the deep learning era \cite{seide2011conversational, sainath2011making, jaitly2012application, DNN4ASR-hinton2012, deng2013recent, yu2017recent}, several attempts were made. The approaches proposed can be divided into two categories: single-channel systems and multi-channel systems, where the difference lies in the number of recording microphones involved. In single-channel systems, the separation process entirely relies on the spectral properties of  speech, such as pitch continuity, harmonic structures, common onsets etc., and this can be achieved by using statistical models~\cite{ephraim1985speech}, rule-based models~\cite{brown1994computational,wang2006computational}, or decomposition-based models~\cite{hershey2010super,fevotte2010notes,mohammadiha2013supervised,chen2013speech}. In multi-channel systems, the separation process can exploit the spatial properties of each source. Various beamforming, or more precisely spatial filtering, methods were proposed by using, for example, Independent Component Analysis~\cite{smaragdis1998blind,sawada2004robust}. Alternatively, clustering-based algorithms attempt to cluster time-frequency bins to individual sources by using spatial features~\cite{conf/icassp/VuH10,sawada2011underdetermined,6638256}. 
There is also a clustering-based approach that uses both spatial and spectral features~\cite{delcroix2011speech}. 
However, regardless of the number of microphones being used, most existing systems work only for rather simple scenarios, e.g. fixed speakers, limited vocabulary, mixtures of different genders etc., and cannot generate satisfying performance for general cases. 

The booming of deep learning has brought progresses in this problem. Different from most other deep learning tasks, multi-talker separation has two unique problems: a permutation problem and an output dimension problem\cite{hershey2016deep,chen2017deep}. The permutation problem arises due to the fact that most deep learning algorithms require estimation targets to be fixed, while in multi-talker separation, arbitrary permutations of the separated sources are equivalent. The output dimension problem refers the fact that the number of mixing speakers varies in different samples, which creates difficulty in learning because a neural network typically requires a fixed dimensionality at its output layer. Three single-channel neural network models were proposed, namely the deep clustering(DC)\cite{hershey2016deep}, deep attractor network(DAN)\cite{chen2017deep} and permutation invariant training(PIT)\cite{yu2017permutation}. In DC and DAN, each time-frequency bin in mixture spectrogram is mapped into a higher dimension representation, i.e. embedding, where the bins from the same speaker are closely located to each other. The two problems are solved by the affinity learning in DC and DAN. PIT was firstly proposed in \cite{hershey2016deep}, and was shown to achieve comparable separation performance in \cite{yu2017permutation}. PIT follows the mask learning framework\cite{wang2014training,chen2017improving}, where the network first generates the output mask for each target speaker, followed by an exhaustive search of combination between the output and the clean reference to fix the permutation problem. The three algorithms largely boost the state of the art in speech separation. The evaluation showed they achieved similar performance for two speaker and three speaker separation on common data sets.

Although the deep learning based methods achieved breakthrough in the cocktail party problem, it is still difficult to apply them to real world applications because of two reasons. Firstly, their separation power has inherent limitation. For example, when there are 4 speakers, even for the most tractable scenario, i.e. two males and two females, single-channel separation seems almost impossible because the mixture is so complex that each speaker's voice is mostly masked by other speakers. 
Secondly, the current single-channel systems are usually vulnerable to reverberation. This would be because the reverberation blurs speech spectral cues which the single-channel separation systems leverage on to isolate each speaker. 

To get rid of the current performance limitation, combining the single-channel and multi-channel approaches would be a natural direction to pursue, since the two approaches use different information for separation and therefore complementary to each other.
 Several neural network based models have been proposed for  multi-channel speech processing, such as acoustic modeling and speech enhancement~\cite{sainath2017multichannel,xiao2016deep,heymann2016neural,xiao2017time}. However, none of the existing systems is able to handle the multi-speaker scenario. For example, in \cite{heymann2016neural,xiao2017time}, a pre-learnt mask is required for each channel, which is impossible for this case simply because there is no existing system to get the mask. In \cite{sainath2017multichannel}, several pooling steps are required, which is not suitable for multi-talker scenario. To the best of our knowledge, currently there is no system that can handle the complex multi-talker speech separation problem.

In this work, we propose a novel, effective yet simple multi-channel speech separation and recognition system. The system consists of a multi-channel part as well as a single-channel part. The multi-channel processing is handled by differential beamformers with 12 fixed beams, equally sampled in space, and followed by the single-channel processing, implemented with anchored deep attractor network\cite{chen2017speaker}, where a ratio mask is learned for each participating source.
By combining the multi- and single-channel processing, the proposed system can fully utilize the spatial and spectral information, and overcome the obstacle inside most multi-channel systems when speakers' location are close, thus leading to better performance than both single- and multi-channel systems. 
The proposed system utilizes the beam as the neural network input, it removes the complex domain processing for neural network, and decouple the spatial and spectral processing, which enables the system to be microphone geometry independent. 
And because of the attractor network architecture, the proposed system performs an end-to-end optimization process, and can be extended to an arbitrary number of sources without the permutation and the output dimension problem.


The rest of the paper is organized as follows. In section \ref{sec:sys} the proposed system is introduced.  Section \ref{sec:exp} describes the experiment setup and the results are discussed in section \ref{sec:dis}. Finally the conclusion is drawn in section \ref{sec:con}
\begin{figure*}[!t]
    \centering
    \includegraphics[width=13.0cm]{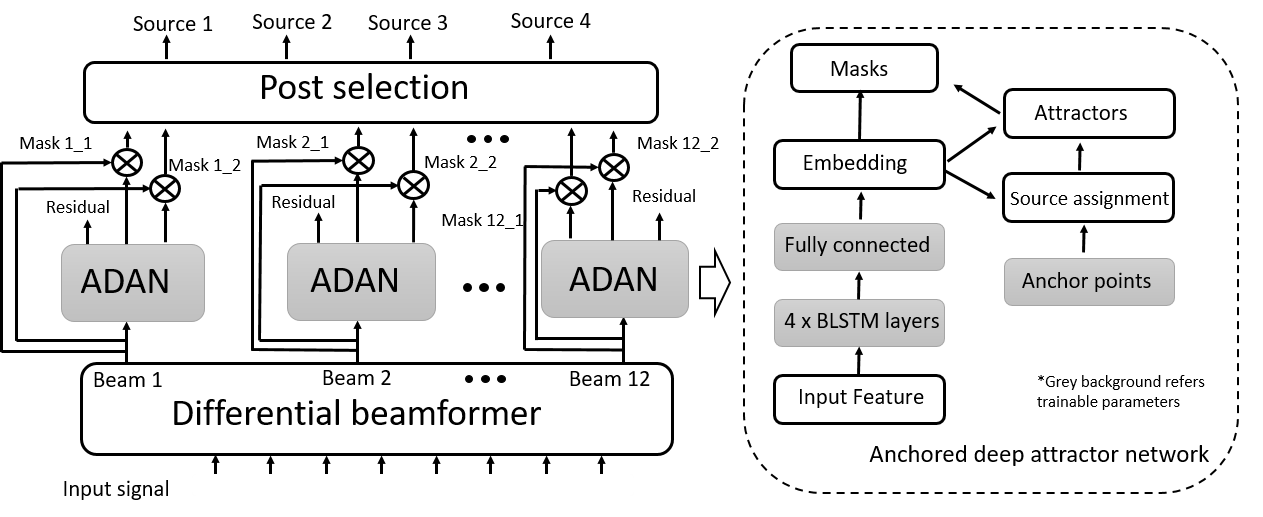}
     \vspace{-0.4cm}
    \caption{\small Proposed system.}
    \label{fig:sys}
\vspace{-0.4cm}
\end{figure*}

\vspace{-0.2cm}
\section{Proposed system}
\label{sec:sys}

A schematic diagram of the proposed system is shown in Fig. \ref{fig:sys}. The proposed system consists of a differential beamformer, which is responsible for multi-channel processing, and a single-channel anchored attractor network, which processes each outputting beam from the beamformer, further separating the beamformed audio. 
The motivation of this architecture is straightforward as follows.
 The single-channel system has inherent limitation in separating complex speaker mixtures. By using beamforming first, the spatial information is utilized to pre-enhance the signal, which reduces the separation complexity, and can be handled by the single-channel system. The multi- and single-channel modules are introduced in Sec. \ref{sec:multi} and Sec. \ref{sec:single}.

\vspace{-0.2cm}
\subsection{Fixed beamformers}
\label{sec:multi}
The multi-channel processing part 
takes multi-channel microphone signals as input and 
provides the subsequent speech separation network with a set of beamformed signals. The beamformer is designed such that each beam has a different look direction. More specifically, 
we uniformly 'sample' the space of direction of arrival with a fixed set of beamformers. Since this is feasible for any microphone arrays, we would expect to obtain a system that is not very sensitive to the geometry of the microphone array to be used.

We propose to utilize differential beamforming to define the beamformer set. Differential beamformers are more attractive in speech separation than additive approaches such as delay-and-sum beamformers. Because differential beaformers can explicitly form acoustic nulls, they can better suppress interfering speakers than the additive beamformers when these speakers are sufficiently spatially isolated from the target speaker direction~\cite{Elko2004,Elko2008}.

In the experiments reported in Section \ref{sec:exp}, we employed a seven-element microphone array. Six microphones were arranged in circle, at the center of which one microphone was placed. Each of the six microphones were separated by 60 degrees. The distance between the center microphone and the other microphones is 42.5 mm. The beamformer is designed following \cite{pessentheinerdifferential}.
In our experiments, we simply used a set of 12 second-order differential beamformers to cover 360$^{\circ}$ degrees. Thus, each acoustic beam was targeted at 0{$^\circ$}, 30{$^\circ$}, and so on. We empirically designed the directivity patterns for the 12 beams. 

In principle, there is a trade-off regarding the number of beams to use. The more beams we have, the more likely one of them is to be targeted at a speaker direction. But, this may complicate the post selection task (see Section \ref{sec:post}).

\vspace{-0.2cm}
\subsection{Anchored deep attractor network}
\label{sec:single}
The single-channel processing of the proposed system is handled by an anchored deep attractor network (ADAN) proposed in \cite{chen2017speaker}, which is a variation of deep attractor network (DAN) . 

Similar to their predecessor deep clustering, an embedding space is formed by neural network in both DAN and ADAN. The separating process in DAN is straightforward: in embedding space, a representative vector for each source, e.g. the average embedding vector, is firstly calculated, which is defined as an attractor point, and serves as the reference for each source.
Then the similarity between the embedding vector and the attractors are calculated for each time-frequency point. This similarity reflects the ``degree of typicalness'' of each time-frequency point, with respect to each source. For example, when an embedding point is close to one attractor, it means the corresponding time-frequency point is more likely to belong to the source represented by that attractor. 
The idea of DAN is to calculate a ratio mask for source separation based on the similarity between the embeddings and the attractors, and the training objective function is to minimize the difference between the masked mixture speech and clean reference.
Compared with DC, DAN allows for end-to-end optimization, where the separation is directly optimized, and thus lead to better performance~\cite{chen2017deep}.

The main difference between DAN and ADAN lies in the attractor forming step. In both DAN and ADAN, the attractor is formed as a weighted  average of the embedding vectors for each source. In DAN, this weight is provided by an oracle binary mask during training. During testing, when the binary mask is not available, DAN adopts a K-means strategy on embedding points to approximate the oracle weights. The problem of this strategy is that a varying degree of mismatch is often observed between training and testing conditions. 
To fix this problem, in ADAN, the oracle mask is not used during training. Instead, an one step K-means with pre-tarined initialization is used in both training and testing.
In detail, the single-channel separation is carried in four steps, as shown in section \ref{sec:4step}


\vspace{-0.1cm}
\subsubsection{Anchored deep attractor network}
\label{sec:4step}
Firstly, as with DC and DAN, an embedding matrix $V\in \mathbb{R}^{T\times F,K}$ is generated by projecting the T-F bins to a high dimensional embedding space by the neural network, as shown in Eqn. \eqref{eqn:emb}, where $T,F,K$ denotes the time, frequency and embedding axes and $\Phi(\cdot)$ refers to the neural network transformation.
\vspace{-0.2cm}
\begin{align}
 V_{tf,k}&= \Phi(X_{t,f})
  \label{eqn:emb}
 \end{align}
Secondly, a pre-segmentation is applied by calculating the distance of the embedding with several pre-trained points in the embedding space, which we refer to as the anchor points. 
More specifically, when there are $C$ mixing sources, we have $N$ anchors denoted as $H \in \mathbb R ^ {N \times K}$. Each source is targeted to bind with one anchor. Then there are  $P = \binom{N}{C}$  total number of source-anchor combinations. For each combination in $P$, we calculate the soft pre-segmentation $W_{p,c,ft}$ by normalizing the distance between each embedding point to each selected anchors, as shown in Eqn. \eqref{eqn:fix_weight}. Based on the soft pre-segmentation $W_{p,c,tf}$, for each combination in $P$,  the attractor $A_{p,c,k}$ is then calculated as the weighted average of embedding points, as shown in eqn. \eqref{eqn:fix_att}.
\vspace{-0.2cm}
\begin{align}
 W_{p, c,tf}&=  \textrm{softmax}\left(\sum_k H_{p,c,k} \times V_{tf,k}\right)
  \label{eqn:fix_weight}
 \end{align}
\vspace{-0.2cm}
where $\textrm{softmax}(\cdot)$ is the softmax activation function over c axis. 
\begin{align}
 A_{p,c,k}&=\frac{\sum_{tf} V_{k,tf}\times W_{p,c,tf} }{\sum_{f,t} W_{p,c,tf}}
 \label{eqn:fix_att}
\end{align} 


After the second step, we obtain $P$ sets of attractors, one in which is used for further processing. In the third step, we compute the in-set similarity defined in Eqn. \eqref{eqn:dis} for each attractor set. The in-set similarity measures the maximum similarity of any of the two attractors within the attractor set. Because the attractors serve as the center for each source to be separated, we select the attractor set that has the minimum in-set similarity. 
\vspace{-0.2cm}
\begin{align}
 S_p&= \max\left\{\sum_k A_{p,i,k} \times A_{p,j,k}\right\}, \quad 1 \leq i < j \leq C
 \label{eqn:dis}
\end{align}
\vspace{-0.2cm}
\begin{align}
 \hat{A}&= \arg\min\{S_p\}, \quad 1 \leq p \leq P
 \label{eqn:select}
\end{align}

In the last step, as with DAN, a mask is formed for each source based on the picked attractor, as shown in Eqn. \eqref{eqn:mask}. The masked speech $X_{f,t} \times M_{f,t,c}$ is compared with the clean references $S_{f,t,c}$ under an squared error measurement in \eqref{eqn:mse}. Unlike the original DAN, the oracle binary mask is not used during the training, therefore the source information is not available at training stage, leading to the permutation problem. We use permutation invariant training~\cite{yu2017permutation} to fix this problem, where the best permutation of the $C$ sources is exhaustively searched. This is done by computing the squared error for each permutation of the masked speech outputs and selecting the minimum error as the error to generate the gradient for back propagation. 
\vspace{-0.2cm}
\begin{align}
M_{f,t,c}&=\textrm{softmax}(\sum_k A_{c,k} \times V_{tf,k} )
\label{eqn:mask}
\end{align}
\vspace{-0.2cm}
\begin{align}
\mathcal{L}&= \sum_{f,t,c}\left \| S_{f,t,c}-X_{f,t} \times M_{f,t,c}\right  \|^2_2
\label{eqn:mse}
\end{align}
\vspace{-0.2cm}

\vspace{-0.2cm}
\subsubsection{Applying ADAN to beamformer outputs}
In the single-channel system, where only one input mixture is available, all the sources are required to be recovered from the mixture in one shot. In contrast, in the proposed multi-beam architecture, multiple beams are available as the input, which gives the single-channel processing more flexibility in measuring the objective thanks to the spatial selectivity of the multi-channel processing. For example, when speakers are adequately separated in space, it is likely each speaker would dominate an individual beam. In this case, the single-channel processing only needs to pick the strongest speaker in each beam and the best separation is guaranteed to exist in the results. If this is the case, the attractor network would reduce to a simple mask learning network. 

However, this assumption is unrealistic. When speakers are close to each other spatially or when there is speakers who are quieter than the others, certain speakers will not be the largest source in any beams. For example, more than 70\% of 4-speaker mixtures in the dataset generated in Section \ref{sec:exp} have such weaker speakers. 
In our single channel processing, for each beam we target $ 1 < G \leq C$ most salient speakers, and add an additional source for the residual signals. More specifically, for each beam, $E=G+1$ output masks are firstly generated from the embedding, among which $G$ sources are selected to compare with the clean references. This process is repeated for all the $\binom{G}{E}$ possibilities. Similarly to the original ADAN, PIT is used where the choice of permutation that leads to the minimum squared error error is selected and that error is used as the final error for back propagation.


In actual processing, when the number of mixing speakers is more than 2, we set $G=2$ and $E=3$, i.e. pick two speakers in each beam and set the rest speakers as one source. Under this strategy, the assumption is that each speaker appears in the first two strongest sources in at least one of the 12 calculated beams. We think that this assumption is not restrictive and would be satisfied in many realistic scenarios if each acoustic beam has a nice directivity pattern.

\vspace{-0.2cm}
\subsection{Post selection}
\label{sec:post}
After performing the single-channel processing for each beamformer output, $E$ outputs will be generated for each of the $B$ beams, resulting in a total of $E \times B$ outputs being produced. Among them, $C$ outputs need to be selected, each corresponding to one of the $C$  target speakers, as the final output. 

The outputs from the single-channel processing can be assumed to be separated to a certain degree, which simplifies the selection.
Several clues may be exploited to accomplish this. For example, we could rely on the assumption that the strongest speaker in one beam is more likely to be dominant in the neighboring beams. Moreover, in actual applications,  speaker locations may also  be detected from non-overlapping  speech segments, which could also help to determine the signals to output. Further investigation is needed for post selection and will be discussed in our follow-up work. In this paper, we used the following simple method.

We firstly calculate an affinity matrix between each of the $E \times B$ magnitude spectrogram by their Pearson correlation.
We then use spectral clustering to group the columns of the affinity matrix into $C+1$ clusters. The motivation for using one additional cluster is to reduce the impacts of failure separation in certain beams and artifacts by isolating them in the additional cluster. 
In our experiments, we observed that after this step, for 92.5\%, 96\% and 99.5\% of 4-speaker, 3-speaker and 2speaker mixtures, each mixing speaker was correctly segregated into different clusters. Therefore, the remaining task after this spectral clustering step is just to select the output in each cluster that has the best speech quality.
Various algorithms can be used for the selection. In this work, we simply use a mean to standard variation criterion $G=\frac{mean}{std}$ of the wavfile to select the best result in each cluster. The determination of noisy cluster is also based on the speech quality.
In this paper, we also report the performance of an oracle selection approach, where we select $C$ outputs that have the best performances in reference to each clean signal. We believe that using a more advanced speech quality estimator would achieve separation performance comparable to that of the oracle selection. 
\vspace{-0.2cm}
\subsection{Qualitative discussion on the proposed architecture}
Compared with previous studies in this field, we think the proposed framework has two main advantages.

Firstly, compared with existing systems that perform only one of either single or multi channel processing, the proposed system exploits both spatial and spectral clues for separation, which has more potential in improving the performance in complex mixing conditions and reverberation.

Secondly, compared with adaptive beamforming, the fixed multi-beam strategy used in the proposed framework would better serve the cocktail party problem because of two reasons. Firstly, in highly complicated mixing scenarios as in the case where there are 4 overlapping speakers, it is not possible with the current technology to obtain parameters required for adaptive beamformering such as steering vectors and noise co-variance matrices. Without accurate estimates of these parameter values, the adaptive beamformer is not able to achieve a reasonable degree of separation. When the adaptive beamformer was used as the multi-channel part of the proposed system, the poor performance in beamforming is propagated to the following single-channel step, possibly leading to further performance degradation. The multi-beam architecture is ensured to generate reasonable separation in certain beams, which will guarantee to have further separation in the single-channel step, i.e. their combination is always beneficial to each other. Secondly, the multi-beam architecture decouples the spatial and spectral processing. When changing the array, as long as we ensure that some of the beams do improve the separation of the speakers, which is generally true unless all speakers are in the same beam, we are guaranteed to get improved performance in the second stage. Adaptive beamformers may produce unstable inputs and not suitable to be used as a preprocessor to the neural networks for single-channel speech separation.

\begin{figure*}[!t]
    \centering
    \includegraphics[width=11cm]{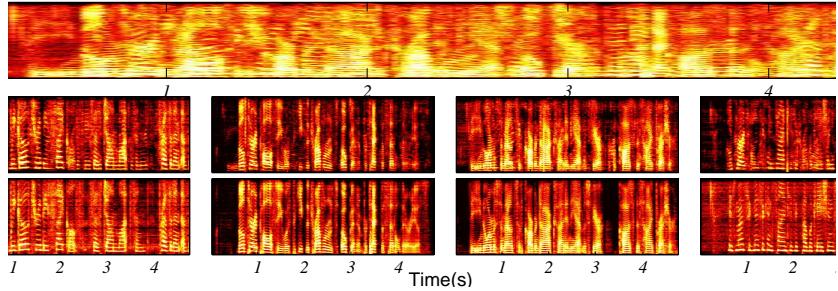}
     \vspace{-0.4cm}
    \caption{\small An example of the mixture spectrogram and the recovered speakers. The example used has similar SDR improvement as the the average performance among all test data.. Upper: original mixture. Middle: recovered utterance. Bottom: the clean reference.}
    \label{fig:res}
\vspace{-0.1cm}
\end{figure*}

\begin{table*}
\small
\centering
\label{tab:sdr_all}
\begin{tabular}{c|c|c|c|c|c|c|c|c|c}
\hline
CLOSED & original & MBBF & MBDAN&OMBDAN & MBIRM &OGEV & OMVDR & IRM & DAN\\
 \hline
2 speakers & -0.25 & +7.28 & +8.95&+11.02 & +15.04 & +1.82 & +11.99 & +10.91 & +7.32\\
3 speakers & -3.2 & +6.86 & +10.20&+11.76 & +14.8 & +4.64 & +12.4& +11.74 & +5.11\\
4 speakers & -4.87 & +6.24 &+9.57 &+11.55 & +13.05 & +6.23 & +11.73 & +12.17 & +4.21\\
\hline
 OPEN & original & MBBF &MBDAN &OMBDAN & MBIRM &OGEV & OMVDR & IRM & DAN\\
\hline
2 speakers & -0.29 & +7.13 &+8.84 &+10.98 & +12.05 & +2.17 & +12.00 & +11.05 & +7.82\\
3 speakers & -3.24 & +7.12 &+9.94 &+11.54 & +15.9  & +4.96 & +12.56 & +11.52 & +5.16\\
4 speakers & -4.89 & +6.37 & +9.53&+11.19 & +12.75 & +6.24 & +11.82 & +12.22 & +4.23\\
\hline
\end{tabular}
\caption{The SDR(dB) improvement for 2,3 and 4 speaker mixture in closed and open speaker set}
\vspace{-0.45cm}
\end{table*}

\vspace{-0.2cm}
\section{Experiment setup}
\label{sec:exp}
\vspace{-0.1cm}
\subsection{Data and pre-processing}
\label{sec:data}
We created a new speech mixture corpus by using utterances taken from the Wall Street Journal (WSJ0) corpus because existing speech separation challenge data sets are not necessarily suitable for evaluation of our model. Three mixing conditions were considered: 2-speaker, 3-speaker, and 4-speaker mixtures. For each condition, a 10-hour training set, a 1-hour validation set, and a 1-hour testing set were created. Similarly to the setting in \cite{chen2017deep}, both our training and validation sets  were sampled from the WSJ0 training set {\footnotesize \verb|si_tr_s|}. Thus, we call the validation set the closed set. Note, however, that there was no utterance appearing in both the training and validation data. The testing data (or the open set) were generated similarly by using utterances of 16 speakers from the WSJ0 development set {\footnotesize \verb|si_dt_05|} and the evaluation set {\footnotesize \verb|si_et_05|}. The testing set had no speaker overlap with the training and validation sets. In each mixture, the first speaker was used as reference, and a mixing SNR was randomly determined for each remaining speaker in the range from -2.5 dB to 2.5 dB. For each mixture, the source signals were truncated to the length of the shortest one to ensure that the mixtures were fully overlapped.

Each multi-channel mixture signal was generated as follows, where we used the image method~\cite{allen1979image} to add reverberation effects. 
Firstly, single-channel clean sources were sampled from the original corpus. 
Then a room with a random size was selected, where a length and a width were sampled from [1m, 10m], a height was sampled from [2.5m, 4m]. An absorption coefficient was also randomly selected from the range of [0.2 0.5]. Positions for a microphone array and each speaker were also randomly determined. For each mixture, we ensured only up to 2 speaker to present within 30$^{\circ}$. Then, a multi-channel room impulse response (RIR)  was generated with the image method, which was  convolved with the clean sources to generate reverberated source signals. They were mixed with randomly chosen SNRs to produce the final mixtue signal to use. 

For speech recognition experiments, we created another testing set by using utterances from WSJ {\footnotesize \verb|si_et_05|}. Here, we used the utterances without pronounced punctuation. Unlike the testing data for speech separation experiments, the source signals were aligned to the longest one by zero padding.  This is because 
signal truncation can be harmful to speech recognition. 
All data were downsampled to 8K Hz. The log spectral magnitude was used as the input feature for a single-channel source separation (i.e., ADAN) network, which was computed by using short-time Fourier transform (STFT) with a 32~ms Hann window shifted by 8 ms. 
\vspace{-0.2cm}
\subsection{Neural network training}
\label{sec:training}
The network used in our experiments consisted of 4 bi-directional long short term memory (BLSTM) layers, each with 300 forward and 300 backward cells. A linear layer with Tanh activation function was added on top of  the BLSTM layers to generate embeddings for individual T-F bins. In the network, 6 anchor points were used, and the embedding space had 20 dimensions. The network was initialized with a single-channel trained 3 speaker separation network which we created in \cite{chen2017speaker} and then finetuned on beamformed signals.  To reduce the training process complexity, for each speaker in each mixture, only the beam where the speaker had the largest SNR was used for training.

For the ASR experiment, we first trained a clean acoustic model which is a 4-layer LSTM-RNN \cite{Sak2014long} with 5976 senones and trained to minimize a frame-level cross-entropy criterion. LSTM-RNNs have been shown to be superior than the feed-forward DNNs \cite{Sak2014long}, which we previously verified with Microsoft Cortana task \cite{Miao16}. Each LSTM layer has 1024 hidden units and the output size of each LSTM layer is reduced to 512 using a linear projection layer. There is no frame stacking, and the output HMM state label is delayed by 5 frames as in \cite{Sak2014long}, with both the singular value decomposition \cite{xue2013restructuring} and frame skipping \cite{Miao16} strategies to reduce the runtime cost. The input feature is the 22-dimension log-filter-bank feature extracted from the 8k sampled speech waveform \cite{Li12mixedband}. The transcribed data used to train this acoustic model comes from 3400 hours of US-English Cortana audio. 

We also built a far-talk ASR model with the domain adaptation method \cite{Li17TS} based on teacher-student (T/S) learning \cite{li2014learning}.  In \cite{Li17TS}, it was shown this T/S learning method is very effective in producing accurate target-domain model.  The far-talk data is simulated from the 3400 hours clean data by using the room impulse response collected from public and Microsoft internal databases. The clean  data  is processed by the clean model (teacher) to generate the corresponding posterior probabilities or soft labels. These posterior probabilities are used in lieu of the usual hard labels derived from the transcriptions to train the target far-talk (student) model with the simulated far-talk data. 
\vspace{-0.3cm}
\subsection{Evaluation}
\label{sec:eval}
Signal-to-distortion ratio (SDR), calculated by bss\_eval tool box\cite{fevotte2005bss_eval}, was used to measure the output speech quality. We report  average SDR improvement as well as SDR improvement for top speakers, i.e. the speakers who have larger mixing SNRs. The ASR performance was evaluated in terms of word error rate (WER). 

To better evaluate the proposed system, six baseline systems were included in our evaluation. We firstly include three oracle beamforming systems: the multi-beam beamformer(MBBF), the oracle generalized eigenvalue (OGEV) beamformer \cite{heymann2016neural} and the oracle minimum variance distortionless response (OMVDR) \cite{van1988beamforming} beamformer. We used the implementation of \cite{heymann2016neural} for OGEV and that of \cite{xiao2017time} for OMVDR. In OGEV, for each speaker, oracle target and noise covariance matrices were calculated from the clean target and the mixture of the rest speakers. In OMVDR, for each speaker, a steering vector was calculated with the known speaker and microphone locations, and oracle noise covariance matrices were calculated from the mixture of the rest speakers. We also tried the oracle MVDR that used only target speaker directions, which is more practical. However, the performance was significantly worse than all other systems, and therefore we do not report its result. In MBBF, the oracle beam selection strategy was adopted. That is, for each speaker, we picked the beam that had largest SDR for that speaker. It should be note that after the MBBF step, most of the beams are still highly mixed because the spatial separation capability of fixed beamformers is limited. Therefore the post selection method described in Sec. 2.3 did not work here. 
We also included the single and multi-channel ideal-ratio-mask (IRM) system for comparison, where a mixture spectrogram was masked by oracle IRMs~\cite{6639038} for each target speaker, and converted to a time domain signal with noisy phase information. As regards the multi-chanel IRM (MBIRM), the multi-channel mixture signal was first processed by the differential beamformer. Then, for each speaker, the IRM approach was applied to the beamformed signal that had the largest SDR for that speaker.  Finally, we also included the single-channel anchored deep attractor network as the baseline, which was trained on the channel 0 of our multi-channel training data.  

\vspace{-0.4cm}
\section{Experiment results}
\label{sec:dis}
\vspace{-0.05cm}

All results are reported in Tables 1--3. Figure \ref{fig:res} shows one separation example, which has a similar SDR gain to the the average SDR improvement over the whole test data.
\vspace{-0.3cm}
\subsection{Speech separation}
\begin{table}[!hbp]
\vspace{-0.3cm}
\small
\centering
\label{tab:sdr_top}
\begin{tabular}{c|c|c|c}
\hline
 Closed & Top 1 & Top 2 & Top 3 \\
 \hline
2 speaker & +11.7 & - & - \\
3 speaker & +13.61 & +11.58 & - \\
4 speaker & +14.86 & +12.56 & +10.39 \\
\hline
Open & Top 1 & Top 1 &Top 3\\
\hline
2 speaker & +11.72 & - & - \\
3 speaker & +13.32 & +11.3 & - \\
4 speaker & +13.71& +11.73 & +10.4 \\
\hline
\end{tabular}
\caption{\small SDR(dB) improvement for selected speaker on OMBDAN}
\end{table}
\vspace{-0.2cm}
\begin{table}[!hbp]
\vspace{-0.2cm}
\small
\centering
\label{tab:asr}
\begin{tabular}{c|c|c|c|c|c}
\hline
Clean model & Mixture & Top 1 & Top 2 & Top3 & Top4\\
\hline
2 speaker & 82.29 & 29.85 & 31.38 & - & -   \\
3 speaker & 93.61 & 31.8 & 39.21 & 44.89& - \\
4 speaker & 95.97  & 42.31 & 46.54 & 53.68 &65.67 \\
\hline
Far-field model & Mixture & Top 1 & Top 2 & Top3 & Top4\\
\hline
2 speaker & 81.96 & 23.6 & 26.38 & - & - \\
3 speaker & 94.19 & 27.95 & 32.64 & 40.61& -\\
4 speaker & 95.91  & 37.79 & 40.29 & 46.1 & 57.93\\
\hline
\end{tabular}
\vspace{-0.2cm}
\caption{\small The WER(\%) for OMBDAN.}
\end{table}
In Table 1, MBDAN refers to the proposed system with spectral clustering-based post selection and OMBDAN refers to the proposed system with oracle selection. While MBDAN underperforms OMBDAN because of the lack of the oracle information, it's performance significantly surpassed that of single-channel DAN. This clearly shows the benefit of multi-channel processing. 
From Fig. 2, we can clearly see differences between the separated sources. We believe improvement in speech quality measurement for post selection would lead to further performance gains. 

Now, we focus on OMBDAN to discuss the separation capability of the model. 
Several observations can be made from Table 1. 
Firstly, the performances for the closed and open speaker sets are very similar, indicating that the system is immune to speaker identity. This is not surprising since the single-channel approaches such as DPCL and DAN also possess the same property. This observation also indicates the potential for further improvement by incorporating the speaker information, which can be available in some real-world scenarios. 
Secondly, compared with other beamforming algorithms (MBBF, OGEV, OMVDR), the proposed framework shows clear advantages. The proposed system with oracle selection achieved over 40\% relative improvement than MBBF and OGEV in all conditions, and achieved similar performance to OMVDR. 
The main difference between OGEV and OMVDR is that the OMVDR has the exact location information, while in OGEV, the steering vector is calculated from the oracle clean covariance matrix, where a small mismatch can be observed because of the reverberation, which might be the reason for the performance difference. In practice, it is almost impossible to obtain those oracle covariances, which makes the proposed system appealing since it only requires the mixing signal as input and can achieve similar performance as oracle MVDR.
Thirdly, both the multi-channel(MBBF) and multi-beam DAN(MBDAN) increase the separation, confirming the fact that they use different information, and are mutually beneficial. Compared with MBIRM, a 1$\sim$3 dB performance loss can be observed in the proposed system. This difference is inline with the output-oracle mismatch in other works such as \cite{chen2015integration,isik2016single}.
Finally, when comparing with the single-channel systems, we can see that the proposed system significantly outperforms the single-channel system. And compared with the result in \cite{chen2017deep,hershey2016deep,yu2017permutation}, where the test set is generated similarly, the single-channel system performs much worse, because of the additional reverberation, which creates the additional ``self-mixing'' in the mixture, and is difficult for single-channel processing. While the multi-channel system removes this problem in beamforming step. And the proposed system achieved similar performance as the single-channel IRM, confirming its efficacy.

Except for the single-channel DAN and MBDAN, all other baseline systems used different forms of oracle information, representing different performance upper-bounds. In this discussion, we used the performance of 4 speaker separation task as example, but the conclusion would be  applicable in other mixing conditions. 
In MBBF, the oracle selection is used, 
which is equivalent to known source location. The performance in MBBF (+6.24dB) roughly sets an upper-bond for all location based beamformers. 
In OGEV and OMVDR, both location and clean source information was given, their performance(+11.73dB) would represent the upper bound that can be achieved with  beamforming or spatial filtering approaches. 
The single-channel IRM uses the clean signal to form the mask, the separation error is mainly introduced by the noisy phase. It is well known that the phase in signal channel processing is difficult to estimate from a single channel mixture, the performance of single channel IRM (+12.17dB), which is usually considered as the upper-bond for all single-channel methods.
And similarly, the MBIRM uses the oracle selection and the IRM information. Its performance (+12.75dB) is the upper-bond of the proposed system.

Table 2 shows the SDR improvement for each separated source with OMBDAN. We can observe that although all sources are improved by a large margin, the best result achieved around 4dB higher than the weaker source. This property is useful in the task where only the dominant speakers are concerned.


\vspace{-0.3cm}
\subsection{Speech recognition}

Table 3 shows the speech recognition performance with separated utterances with OMBDAN, with the clean trained and far field acoustic model. In table 3, in all conditions, the WER of the mixture speech is close to 100\%. After processing, the WER decreased by a large margin in all conditions, leading 62.80\%, 58.73\%, 45.59\% relative gain with clean model and 69.51\%, 64.19\%, 52.53\% with far-field model.
The far field model achieved better performance because the reverberation and stationary noise is included in the training data. 

It should be noted that both clean and far field AM are trained with only unmixed signals. With retraining on the separated speech, a much better performance could be expected. Moreover, a even better performance can be achieved by directly optimizing the ASR performance, rather than using the separation-recognition scheme.

\vspace{-0.3cm}
\section{Future work and conclusion}
\vspace{-0.1cm}
\label{sec:con}

In the paper, we present a multi-channel speech separation system, which separated the mixture speech through a set of fixed beamformers, followed by a single-channel anchored deep attractor network. The proposed system combines the both the spatial and spectral information and largely increases the state-of-the-art performance in the cocktail party problem in speech separation and recognition, and can handle very challenging senarios, e.g. 4 same gender speakers.

In cocktail party problem, there are two branches, the speech separation and the speech recognition. In single-channel systems, the two tasks are similar. One can easily shift to another by simply changing the objective and training target, e.g. from the reconstructing error to cross entropy. However, in multi-channel processing, there are clear differences in the two tasks. This is because in separation, a mask is usually learned for each speaker, however the beamformed signals to be masked varies for each speaker. Therefore the separation process must be independent for each speaker. While in ASR, since the target is text information(e.g. senone, characters etc.), which remains the same for all beamformed signals, the recognition system in the same scenario could be established by jointly optimizing all the beams, and directly outputting the posterior for all speakers, which removes the post selection step. Since this optimization is performed end-to-end, the recognition error is expected to be much lower. The exploration of this direction will be described in a follow-up work.

\vfill\pagebreak

\bibliographystyle{IEEEbib}
\bibliography{strings,refs}

\end{document}